\documentclass{sf2a-conf2024}
\usepackage{graphicx}
\usepackage{hyperref}
\usepackage[]{natbib}  
\usepackage{epstopdf}
\usepackage{xcolor}

\def\BibTeX{{\rm B\kern-.05em{\sc i\kern-.025em b}\kern-.08em
    T\kern-.1667em\lower.7ex\hbox{E}\kern-.125emX}}
\bibpunct{(}{)}{;}{a}{}{,}  

\begin{document}

\TitreGlobal{SF2A 2024}


\title{Improving the parametrization of transport and mixing processes in planetary atmospheres:\\
 the importance of implementing the full Coriolis acceleration}

\runningtitle{ }

\author{C. Moisset}\address{Université Paris-Saclay, Université Paris Cité, CEA, CNRS, 91191 Gif-sur-Yvette, France}
\author{S. Mathis$^1$}
\author{P. Billant}\address{LadHyX, Ecole Polytechnique / CNRS, Palaiseau, France}
\author{J. Park}\address{Centre for Fluid and Complex Systems, Coventry University, Priory Street, Coventry CV15FB, UK}

\setcounter{page}{237}


\maketitle


\begin{abstract}
With the ongoing characterisation of the atmospheres of exoplanets by the JWST, we are unveiling a large diversity of planetary atmospheres, both in terms of composition and dynamics. As such, it is necessary to build coherent atmospheric models for exoplanetary atmospheres to study their dynamics in any regime of thickness, stratification and rotation. However, many models only partially include the Coriolis acceleration with only taking into account the local projection of the rotation vector along the vertical direction (this is the so-called “Traditional Approximation of Rotation") and do not accurately model the effects of the rotation when it dominates the stratification.\\

In this contribution, we report the ongoing efforts to take the full Coriolis acceleration into account for the transport of momentum and the mixing of chemicals. First, we show how the horizontal local component of the rotation vector can deeply modifies the instabilities of horizontal sheared flows and the turbulence they can trigger. Next, we show how the interaction between waves and zonal winds can be drastically modified because of the modification of the wave damping or breaking when taking into account the full Coriolis acceleration. These works are devoted to improve the parameterization of waves and turbulent processes in global atmospheric models.
\end{abstract}



\section{Introduction}
  Since the first successful exoplanet detection in 1995 \citep{MayorQueloz1995}, the number of planets identified outside of our solar system has been growing, to reach more than thousands today. Their detection is accompanied by the characterisation of their atmospheres, a task first undertaken by the Spitzer and Hubble space telescopes, that the JWST is now bringing to a unprecedented level of precision \citep[e.g.][]{Dyreketal2024, Belletal2024}. As our solar system on its own already harbors an  impressive diversity of planetary atmospheres (see tab. \ref{tab_atm}), we expect these efforts to unveil a significant number of new possible atmospheric configurations in the coming years.\\
\begin{table}[ht!]
  \begin{center}
  \resizebox{0.6\textwidth}{!}{
  \begin{tabular}{lcccc}
      {\bf Atmospheres}  & {\bf Earth} & {\bf Venus}  & {\bf Jupiter} &  {\bf Titan} \\[3pt]
       Chemical Components  & N$_{2}$, O$_{2}$ & CO$_{2}$, N  & H$_{2}$, He & N$_{2}$, CH$_{4}$ \\
       Thickness [km]   & $100$ & $300$ & $5000$ & $200$ \\
       Brunt-V\"ais\"al\"a frequency, $N_{max}$ [s$^{-1}$]   & $3\times10^{-2}$ & $3\times10^{-2}$ &  $2\times10^{-2}$ & $4\times10^{-3}$ \\
       Rotation rate, $2\Omega$ [rad.s$^{-1}$]   & $1.4\times10^{-4}$ & $3.6\times10^{-5}$  & $3.5\times10^{-4}$ & $1\times10^{-5}$ \\
       $N_{\rm max}/(2\Omega)$ & $214$ & $834$  & $57$ & $400$
  \end{tabular}
  }
  \caption{Characteristics of a few planetary atmospheres in our solar system: main chemical composition, thickness, maximum Brunt-V\"ais\"al\"a frequency, rotation rate and their ratio $N_{\rm max}/(2\Omega)$. $N/(2\Omega)$ ranges from $0$ to $N_{\rm max}/(2\Omega)$.}
  \label{tab_atm}
  \end{center}
\end{table}
Understanding the atmospheric dynamics of (exo)planets is key for our interpretation of satellite observations, like that of radiative thermal emissions or of abundances of chemical elements, and to determine their climate and habitability \citep[e.g.][]{KodamaTurbet2024, Teinturieretal2024}. To do so, we use Global Circulation Models (GCM) as they allow a coherent treatment of all physical processes occurring in these atmospheres. In this framework, the large-scale dynamics (like the zonal flows, the circulation cells, the pressure and temperature profiles,...) is well resolved by GCMs but the smaller, sub-grid processes (the dynamics of waves, instabilities, turbulence, convection,...) need to be parametrized.\\
\indent Planetary atmospheres can be modelled as rotating layers of stratified fluid but current GCM often use the Traditional Approximation of the rotation (TA), where the rotation is only partially implemented through its projection on the local vertical direction \citep[see the discussion in][]{TortDubos2014}. This approach is only valid in the case where the buoyancy force strongly dominates the Coriolis acceleration along the vertical direction. \cite{GerkemaZimmerman2008} have however shown how the TA overly simplifies the dynamics in some cases, especially when the rotation modifies the dynamics as significantly as the stratification. Considering the large diversity of planetary atmospheres that we expect to uncover, our dynamical models of atmospheres need to be robust enough to adapt to any regime of stratification, rotation and thickness. Parametrizations, in particular, need to include the full Coriolis acceleration to be able to tackle any value of the ratio $N$/2$\Omega$, where $N$, the Brunt-V\"ais\"al\"a frequency measures the intensity of the stratification and 2$\Omega$, the rotation rate, that of the rotation. Investigations in this direction are beginning to appear \citep[e.g.][]{GerkemaShrira2005, Zeitlin2018, Parketal2021, ToghraeiBillant2022}. Our work focuses on providing prescriptions for the transport and mixing for any $N$/2$\Omega$: we provide a new understanding of the instability of horizontally sheared zonal flows and new parametrizations for the damping and breaking of inertia-gravity waves.
\section{Non-traditional and non-linear dynamics of the instability of horizontally sheared zonal flows}
We study the impact of the full Coriolis acceleration over the non-linear horizontal shear instability exploring different values of the stratification, via the buoyancy frequency $N$, and the non-traditional Coriolis parameter $\tilde f=2\Omega\cos(\phi)$, where  $\phi$ is the latitude. We use the pseudo-spectral code NS3D \citep{Deloncleetal2014} to monitor, via DNS, the growth and development of the instability of a zonal flow sheared with an hyperbolic tangent latitudinal profile in a local Cartesian horizontal plane \citep[e.g.][]{Parketal2021}. The code solves the 2D Navier-Stokes equations using the Boussinesq approximation.
The Reynolds number is fixed to $Re=2000$ with the Prandtl number $Pr=\nu/\kappa=1$, where $\nu$ and $\kappa$ are the kinematic viscosity and thermal diffusivity, respectively. The impact of the Reynolds and Prandtl numbers will be explored in future studies.\\
\begin{figure}[ht!]
 \centering
 \includegraphics[width=0.75\textwidth,clip]{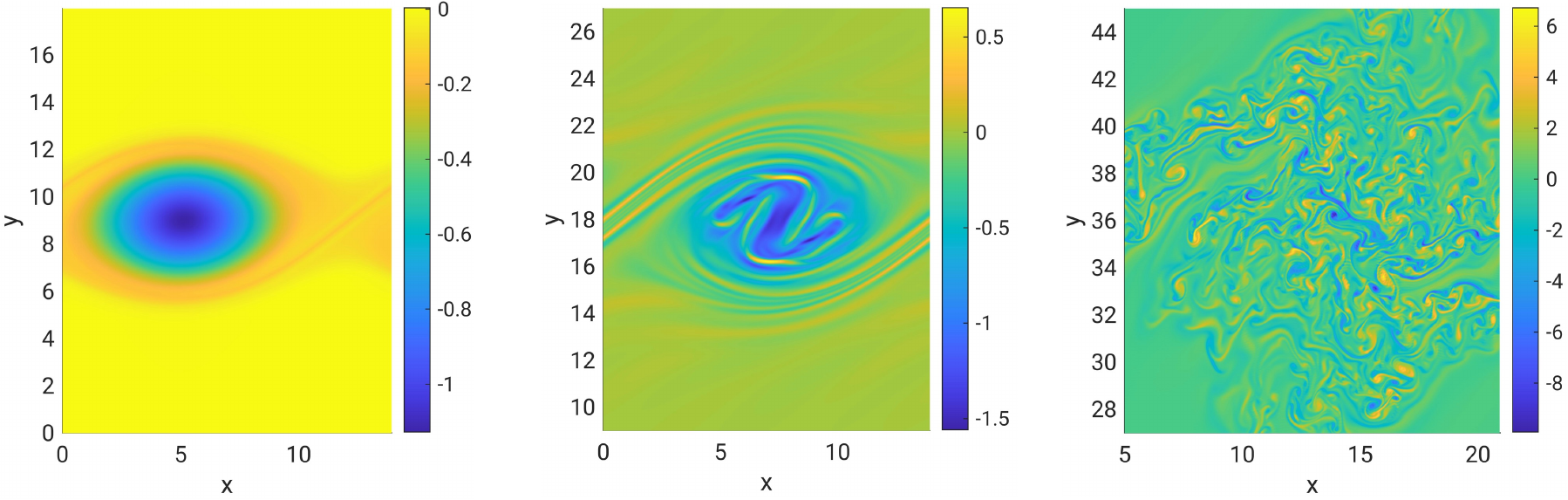}
  \caption{Field of 2D vertical vorticity (with $x$ the zonal and $y$ the latitudinal coordinates) for different values of $N$ and $\tilde{f}$: {\bf Left:} Strongly stratified regime ($N$, $\tilde{f}$)=(5, 1). {\bf Middle:} ($N$, $\tilde{f}$)=(1, 0.5). {\bf Right:} Weakly stratified regime ($N$, $\tilde{f}$)=(0.5, 0.5). Taken from Moisset et al., in prep.}
  \label{Shear Inst}
\end{figure}
We identified three different scenarios of non-linear evolution of the instability (see fig. \ref{Shear Inst}). When the stratification largely dominates the rotation (fig. \ref{Shear Inst}, left panel): the flow evolves towards a stable and  quasi-stationary vortex as observed when using the Traditional Approximation. For a weaker stratification, the vortex still forms but is perturbed by localised secondary instabilities (fig. \ref{Shear Inst}, middle panel). Then, when rotation is comparable to the stratification, the vortex is subjected to very strong secondary instabilities and the flow becomes fully turbulent (fig. \ref{Shear Inst}, right panel).\\
As such, stratification and rotation have competing effects: the stratification favors the “stable” vortex regime while the non-traditional effects act to destroy that structure via secondary instabilities. We observed that, non-traditional effects stretch the vortex along the latitudinal direction when the non-linear dynamics sets in. These stretched vorticity braids become thinner and new inflection points appear : secondary instabilities are then more likely to develop and lead to turbulence.\\
Therefore, we show that the non-traditional part of the Coriolis acceleration leads to a strongly different non linear evolution of the shear instability, that should impact the transport of heat and chemicals: when the full rotation competes with the stratification, it leads to turbulent mixing that would be completely missed by prescriptions within the framework of the Traditional Approximation.
\section{Non-traditional inertia-gravity waves damping and non-linear breaking} 
In planetary atmospheres, viscous and heat diffusion or convective breaking are two channels through which inertia-gravity waves are able to locally deposit angular momentum \citep[e.g.][]{LottGuez2013}. These processes are at the heart of atmospheric wave-mean flow interactions and are able to significantly alter the wind field and general circulation \citep[e.g.][]{Liuetal2023}. In this work, we follow \cite{VadasFritts2005} and we assume that the momentum and heat diffusion are first-order perturbations. We take the full Coriolis acceleration into account following \cite{GerkemaShrira2005} and \cite{MathisNeinerTranMihn2014}.\\
\begin{figure}[ht!]
\centering
\begin{minipage}[b]{0.45\linewidth}
 \centering
 \includegraphics[width=\linewidth,clip]{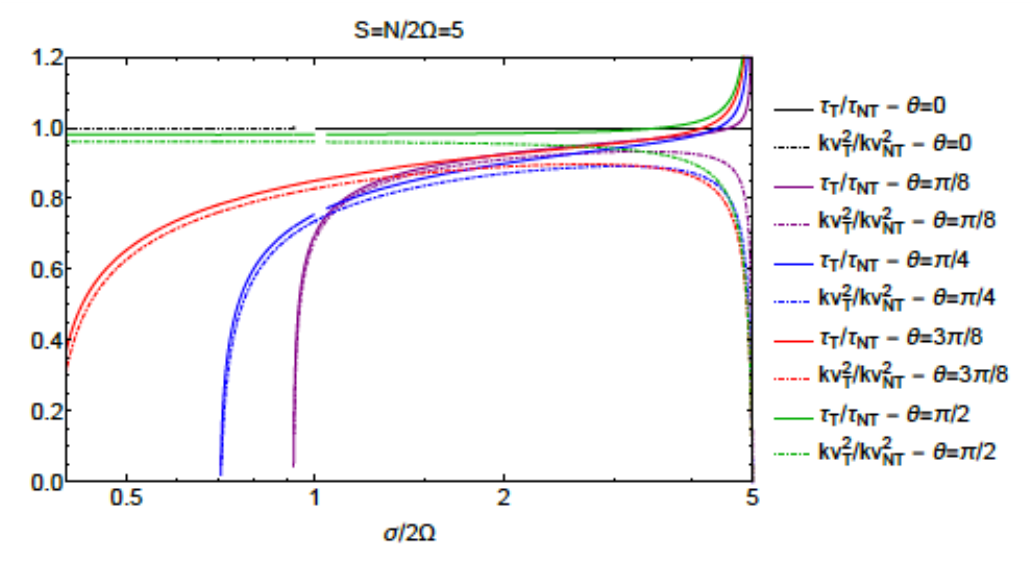}
\end{minipage}
\begin{minipage}[b]{0.5\linewidth}
\centering
  \caption{Comparison of the predicted damping with and without TA ($\tau_\textrm{T}$ and $\tau_{\textrm{NT}}$ respectively) as a function of the normalised frequency for different given co-latitudes ($0$ being the pole and $\pi/2$ the equator) for a fixed ratio $N/2\Omega=5$. Similar comparison for the squared vertical wavenumber squared (kv$^2_\textrm{T}$ and kv$^2_{\textrm{NT}}$) are also provided (Mathis, in prep.).}
  \label{Damping_IGW}
  \vspace{0.9 cm}
\end{minipage}
\end{figure}
Figure \ref{Damping_IGW} clearly shows how the TA drastically underestimates the vertical wave number (i.e. underestimates the vertical wavelength) and the damping for inertia gravity waves of frequency close to and below $2\Omega$ (which is the so-called "sub-inertial regime", cf. \cite{GerkemaZimmerman2008}). As a consequence, we predict with the TA a deposit of the momentum by inertia-gravity waves farther away from their excitation region than when taking the full Coriolis acceleration into account. As a consequence, with TA, prescriptions regarding the vertical deposit of momentum by waves expect the flux of momentum to penetrates higher in the atmosphere than what we found when the full rotation is implemented. As a result, should the wave interact with mean flows, it would be at lower altitude than what we found using the TA. It becomes relevant then, to implement a non traditional parametrisation in GCM (Mathis, in prep.).\\

For the model of the breaking of inertia-gravity waves, we follow the method of the work of \cite{Lindzen1981} and \cite{LottGuez2013} to determine the threshold at which the convective instability that generates the breaking sets up. From this, we derive a saturation condition on the velocity field that we use to compute the transported flux of angular momentum.\\
\begin{figure}[ht!]
 \centering
 \includegraphics[width=0.78\textwidth,clip]{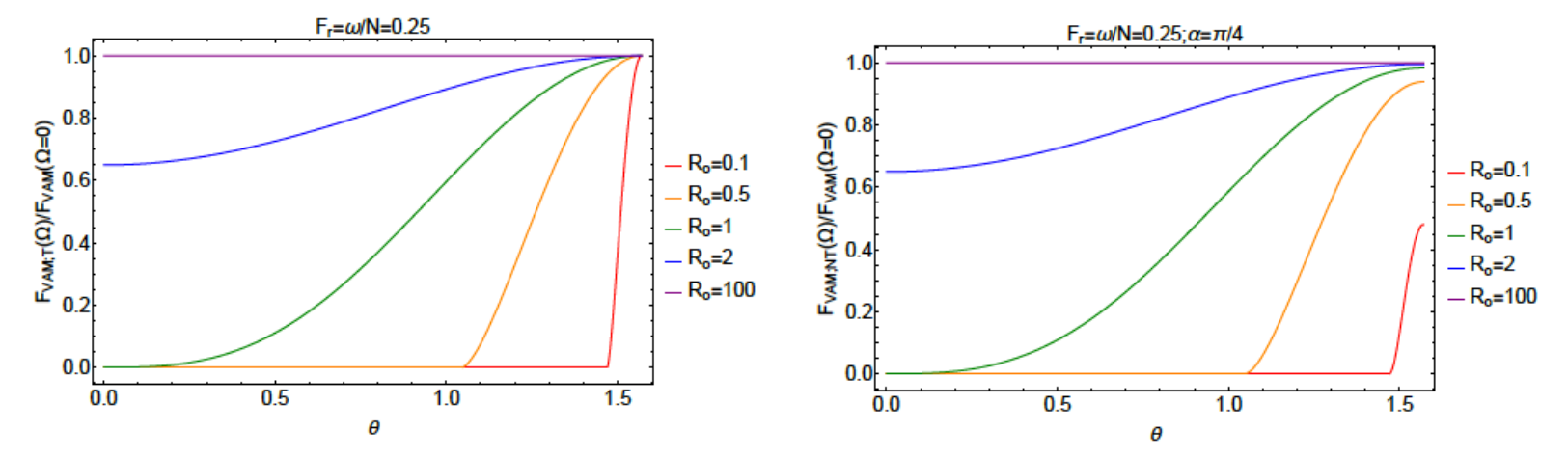}
  \caption{{\bf Left:} Vertical flux of angular momentum computed using TA compared with the non-rotating case, for a given Froude number ($\omega/N$, where $\omega$ is the frequency of the wave), as a function of the co-latitude $\theta$. Profiles are given for different waves' Rossby numbers Ro=$\omega/2\Omega$ (Ro$<1$ being the sub-inertial regime). {\bf Right:} Same as in the left but the flux is computed with the full Coriolis acceleration and compared with the non-rotating case (Mathis, in prep.).}
  \label{Breaking IGW}
\end{figure}
Figure \ref{Breaking IGW} shows how the flux of angular momentum caused by the breaking of inertia gravity waves are overestimated by the TA. Indeed, for waves of frequency comparable to the rotation, in particular at the equator, the TA prescribes that the action of the Coriolis acceleration vanishes. It then underestimates the action of the rotation which, among other mechanisms, stabilize the convective instability that leads to the breaking. As such, the TA prescribes an effect of the breaking that is more important than what we predict it to be when implementing the complete rotation.\\
For the dynamics of an atmosphere, it means that the deposition of angular momentum by sub-inertial waves around the Equator is likely to be overestimated in current prescription, which would lead to interactions with mean flows and circulations that may be weaker in the reality (Mathis, in prep.).
\section{Conclusions}
As we anticipate the growing number of exoplanet detection and the characterization of their atmosphere, we highlight the need of general and robust models of their dynamics for any rotation, stratification and thickness. In particular, we examine the key differences in the transport of momentum by the instabilities of latitudinally sheared flows and by the damping and breaking of inertia gravity waves when the full Coriolis acceleration is implemented.\\
We have showed that the TA can mask the turbulence and mixing occurring due to the shear instability when the rotation becomes comparable to the stratification as for instance in equatorial regions and in weakly stably stratified layers.\\
Concerning inertia-gravity waves, we have demonstrated that the TA underestimates the effect of the damping but also tends to overestimate that of the breaking. This can drastically affect the strength of wave-mean flows interactions and the altitude at which they occur, especially in the sub-inertial regime. For both mechanisms, it would impact the interaction between these waves and mean flows, especially in the lower range of allowed frequencies.\\
All of this demonstrates how it is necessary to go beyond the traditional approximation when modeling and simulating the dynamics of planetary atmospheres. In both cases, the difference brought by the full implementation of the rotation is bond to modify the dynamics of the transport processes and consequently the structures of the modeled atmospheres.\\
\begin{acknowledgements}
C.M. and S.M. acknowledge support from the European Research Council (ERC) under the Horizon Europe program (Synergy Grant agreement 101071505: 4D-STAR), from the CNES SOHO-GOLF and PLATO grants at CEA-DAp, and from PNP and PNPS (CNRS/INSU). While partially funded by the European Union, views and opinions expressed are however those of the authors only and do not necessarily reflect those of the European Union or the European Research Council. Neither the European Union nor the granting authority can be held responsible for them.
\end{acknowledgements}
\bibliographystyle{aa}  
\bibliography{Moisset_S13.bib} 

\begin{thebibliography}{17}
\expandafter\ifx\csname natexlab\endcsname\relax\def\natexlab#1{#1}\fi

\bibitem[{{Bell} {et~al.}(2024){Bell}, {Crouzet}, {Cubillos}, {Kreidberg}, {Piette}, {Roman}, {Barstow}, {Blecic}, {Carone}, {Coulombe}, {Ducrot}, {Hammond}, {Mendon{\c{c}}a}, {Moses}, {Parmentier}, {Stevenson}, {Teinturier}, {Zhang}, {Batalha}, {Bean}, {Benneke}, {Charnay}, {Chubb}, {Demory}, {Gao}, {Lee}, {L{\'o}pez-Morales}, {Morello}, {Rauscher}, {Sing}, {Tan}, {Venot}, {Wakeford}, {Aggarwal}, {Ahrer}, {Alam}, {Baeyens}, {Barrado}, {Caceres}, {Carter}, {Casewell}, {Challener}, {Crossfield}, {Decin}, {D{\'e}sert}, {Dobbs-Dixon}, {Dyrek}, {Espinoza}, {Feinstein}, {Gibson}, {Harrington}, {Helling}, {Hu}, {Iro}, {Kempton}, {Kendrew}, {Komacek}, {Krick}, {Lagage}, {Leconte}, {Lendl}, {Lewis}, {Lothringer}, {Malsky}, {Mancini}, {Mansfield}, {Mayne}, {Evans-Soma}, {Molaverdikhani}, {Nikolov}, {Nixon}, {Palle}, {Petit dit de la Roche}, {Piaulet}, {Powell}, {Rackham}, {Schneider}, {Steinrueck}, {Taylor}, {Welbanks}, {Yurchenko}, {Zhang}, \& {Zieba}}]{Belletal2024}
{Bell}, T.~J., {Crouzet}, N., {Cubillos}, P.~E., {et~al.} 2024, Nature Astronomy, 8, 879

\bibitem[{{Deloncle}(2014)}]{Deloncleetal2014}
{Deloncle}, A. 2014, Ecole Polytechnique

\bibitem[{{Dyrek} {et~al.}(2024){Dyrek}, {Min}, {Decin}, {Bouwman}, {Crouzet}, {Molli{\`e}re}, {Lagage}, {Konings}, {Tremblin}, {G{\"u}del}, {Pye}, {Waters}, {Henning}, {Vandenbussche}, {Ardevol Martinez}, {Argyriou}, {Ducrot}, {Heinke}, {van Looveren}, {Absil}, {Barrado}, {Baudoz}, {Boccaletti}, {Cossou}, {Coulais}, {Edwards}, {Gastaud}, {Glasse}, {Glauser}, {Greene}, {Kendrew}, {Krause}, {Lahuis}, {Mueller}, {Olofsson}, {Patapis}, {Rouan}, {Royer}, {Scheithauer}, {Waldmann}, {Whiteford}, {Colina}, {van Dishoeck}, {{\"O}stlin}, {Ray}, \& {Wright}}]{Dyreketal2024}
{Dyrek}, A., {Min}, M., {Decin}, L., {et~al.} 2024, \nat, 625, 51

\bibitem[{{Gerkema} \& {Shrira}(2005)}]{GerkemaShrira2005}
{Gerkema}, T. \& {Shrira}, V.~I. 2005, J. Fluid Mech., 529, 195

\bibitem[{{Gerkema} {et~al.}(2008){Gerkema}, {Zimmerman}, {Maas}, \& {van Haren}}]{GerkemaZimmerman2008}
{Gerkema}, T., {Zimmerman}, J.~T.~F., {Maas}, L.~R.~M., \& {van Haren}, H. 2008, Reviews of Geophysics, 46, RG2004

\bibitem[{{Kodama} \& {Turbet}(2024)}]{KodamaTurbet2024}
{Kodama}, T. \& {Turbet}, M. 2024, in AAS/Division for Extreme Solar Systems Abstracts, Vol.~56, AAS/Division for Extreme Solar Systems Abstracts, 629.05

\bibitem[{{Lindzen}(1981)}]{Lindzen1981}
{Lindzen}, R.~S. 1981, \jgr, 86, 9707

\bibitem[{{Liu} {et~al.}(2023){Liu}, {Millour}, {Forget}, {Gilli}, {Lott}, {Bardet}, {Gonz{\'a}lez Galindo}, {Bierjon}, {Naar}, {Martinez}, {Lebonnois}, {Fan}, {Pierron}, \& {Vandemeulebrouck}}]{Liuetal2023}
{Liu}, J., {Millour}, E., {Forget}, F., {et~al.} 2023, Journal of Geophysical Research (Planets), 128, e2023JE007769

\bibitem[{{Lott} \& {Guez}(2013)}]{LottGuez2013}
{Lott}, F. \& {Guez}, L. 2013, Journal of Geophysical Research (Atmospheres), 118, 8897

\bibitem[{{Mathis} {et~al.}(2014){Mathis}, {Neiner}, \& {Tran Minh}}]{MathisNeinerTranMihn2014}
{Mathis}, S., {Neiner}, C., \& {Tran Minh}, N. 2014, \aap, 565, A47

\bibitem[{{Mayor} \& {Queloz}(1995)}]{MayorQueloz1995}
{Mayor}, M. \& {Queloz}, D. 1995, \nat, 378, 355

\bibitem[{{Park} {et~al.}(2021){Park}, {Prat}, {Mathis}, \& {Bugnet}}]{Parketal2021}
{Park}, J., {Prat}, V., {Mathis}, S., \& {Bugnet}, L. 2021, \aap, 646, A64

\bibitem[{{Teinturier} {et~al.}(2024){Teinturier}, {Charnay}, {Spiga}, {B{\'e}zard}, {Leconte}, {Mechineau}, {Ducrot}, {Millour}, \& {Cl{\'e}ment}}]{Teinturieretal2024}
{Teinturier}, L., {Charnay}, B., {Spiga}, A., {et~al.} 2024, \aap, 683, A231

\bibitem[{{Toghraei} \& {Billant}(2022)}]{ToghraeiBillant2022}
{Toghraei}, I. \& {Billant}, P. 2022, J. Fluid Mech., 950, A29

\bibitem[{{Tort} \& {Dubos}(2014)}]{TortDubos2014}
{Tort}, M. \& {Dubos}, T. 2014, Quarterly Journal of the Royal Meteorological Society, 140, 2388

\bibitem[{{Vadas} \& {Fritts}(2005)}]{VadasFritts2005}
{Vadas}, S.~L. \& {Fritts}, D.~C. 2005, Journal of Geophysical Research (Atmospheres), 110, D15103

\bibitem[{{Zeitlin}(2018)}]{Zeitlin2018}
{Zeitlin}, V. 2018, Phys. Fluids, 30, 061701

\end{thebibliography}

\end{document}